\documentclass[12pt,amsmath]{iopart}

\usepackage{graphicx}
\usepackage{iopams}

\newcommand{\cmsec}{{\mathrm{cm}^{-2}~\mathrm{sec}^{-1}}}
\newcommand{\cms}{{\mathrm{cm}}}
\newcommand{\second}{{\mathrm{sec}}}
\newcommand{\GeV}{\mathrm{GeV}}
\newcommand{\pcs}{\mathrm{pc}}
\newcommand{\sr}{\mathrm{sr}}
\newcommand{\Mmin}{M_{\mathrm{min}}}
\newcommand{\Mmax}{M_{\mathrm{max}}}
\newcommand{\Lmin}{L_{\mathrm{min}}}
\newcommand{\Lmax}{L_{\mathrm{max}}}

\def\VEV#1{\langle #1 \rangle}

\begin{document}

\title[The Gamma-Ray-Flux PDF
    from Galactic Halo Substructure]{The Gamma-Ray-Flux Probability Distribution Function
    from Galactic Halo Substructure}

\author{Samuel K.~Lee, Shin'ichiro Ando, and Marc Kamionkowski}
\address{California Institute of Technology, Mail Code 350-17,
Pasadena, CA 91125}
\ead{sklee@astro.caltech.edu, ando@tapir.caltech.edu, kamion@tapir.caltech.edu}

\date{\today}

\begin{abstract}
One of the targets of the recently launched Fermi Gamma-ray Space Telescope is a diffuse gamma-ray background from dark-matter annihilation or decay in the Galactic halo.  N-body simulations and theoretical arguments suggest that the dark matter in the Galactic halo may be clumped into substructure, rather than smoothly distributed.  Here we propose the gamma-ray-flux probability distribution function (PDF) as a probe of substructure in the Galactic halo.  We calculate this PDF for a phenomenological model of halo substructure and determine the regions of the substructure parameter space in which the PDF may be distinguished from the PDF for a smooth distribution of dark matter.  In principle, the PDF allows a statistical detection of substructure, even if individual halos cannot be detected.  It may also allow detection of substructure on the smallest microhalo mass scales, $\sim M_\oplus$, for weakly-interacting massive particles (WIMPs).  Furthermore, it may also provide a method to measure the substructure mass function.  However, an analysis that assumes a typical halo substructure model and a conservative estimate of the diffuse background suggests that the substructure PDF may not be detectable in the lifespan of Fermi in the specific case that the WIMP is a neutralino.  Nevertheless, for a large range of substructure, WIMP annihilation, and diffuse background models, PDF analysis may provide a clear signature of substructure.
\end{abstract}


\maketitle

\section{Introduction} \label{sec:intro}

It has long been a goal of astrophysics and cosmology to
determine the distribution and nature of the dark matter that
populates our Galactic halo.  Only more recently have we begun to focus on the possibility to detect substructures in the Galactic halo \cite{substructure}.  In hierarchical structure formation,
small gravitationally bound dark-matter systems form first and
then merge to form progressively more massive systems.
In each stage, some of the earlier generations of structure may remain intact after merging, and so the Milky Way halo may contain substructures
over a wide array of masses. Scaling arguments suggest that
substructures may continue all the way down to the smallest
mass scales at which there is primordial power
\cite{Kamionkowski:2008vw,earthmass}, although the precise details may be
uncertain \cite{hsz}.  If
weakly-interacting massive particles (WIMPs)
\cite{Jungman:1995df} make up the dark matter,
the cutoff mass should be in the range
$10^{-4}-10^{3}\,M_\oplus$ \cite{Profumo:2006bv}, and if axions
\cite{axionreviews}
make up the dark matter, it may be as small as
$10^{-12}\,M_\oplus$ \cite{Johnson:2008se}.

If WIMPs make up the dark matter, there may be several avenues
toward detecting them.
With the launch of the Fermi Gamma-ray Space Telescope (formerly GLAST) \cite{glast}, however, there
is now particular attention being paid to detection of energetic
gamma rays from dark-matter annihilation in the Galactic halo
(see, e.g., Ref.~\cite{Baltz:2008wd} and references therein).
While the diffuse flux from such annihilations have been
considered for a long time \cite{subgamma}, the possibility to detect
substructure, through angular variations in the background, is
more recent \cite{anisotropy1, anisotropy2, anisotropy3}.
It is possible that individual substructures may be
resolved \cite{Hooper:2007be}.  Proper motions of the smallest
microhalos have also been considered \cite{proper1, proper2}.

In this paper, we propose the one-point gamma-ray-flux probability
distribution function (PDF) as a probe of halo substructure.  If dark
matter is smoothly distributed, then the variation in the number
of diffuse-background photons from one pixel to another should
arise only from Poisson fluctuations.  If, however, there is
substructure, there will be additional flux variations from
pixel to pixel.  This may provide another route---an alternative
to the angular two-point correlation function
\cite{anisotropy1, anisotropy3}---to detect
substructure statistically, especially for the very smallest
microhalo mass scales.\footnote{It has been similarly suggested
\cite{Willis:1996} that background fluctuations may be used to
learn about the traditional astrophysical sources contributing to the
diffuse background.}  It may also allow measurement of the substructure mass function, under certain model assumptions outlined below.

We illustrate with a phenomenological model for Galactic substructure
in which a fraction $f$ of the halo is made of dark-matter
microhalos with a power-law mass function (with a lower mass
cutoff $\Mmin$) and a constant mass--to--gamma-ray-luminosity
ratio $\Upsilon=\Mmin/\Lmin$.  The next Section introduces
this model and discusses the constraints from the Energetic
Gamma Ray Experiment Telescope (EGRET) \cite{egret} to the parameter
space.  In Section \ref{sec:onepoint}, we calculate the flux PDF
for this model and discuss the translation to a discrete
distribution of counts in each Fermi pixel.  We provide in
Section \ref{sec:genfeat} numerical results for the PDF for
an illustrative model.  Section \ref{sec:detect}
determines the regions of the parameter
space in which the PDF of substructure can be distinguished
from that of a smoothly distributed background.  In Section
\ref{sec:conclusions} we summarize and comment on additional
steps that must be taken to implement this probe.

\section{Substructure/Annihilation Models and EGRET Constraints} \label{sec:egret}

\subsection{Halo Model and Microhalo Mass Function}
\label{sec:halomodel}

We assume that a fraction $f$ of the dark matter in the
Galactic halo is composed of objects with a power-law mass
function $dn_h / dM_h \propto M_h^{-\alpha}$,
independent of Galactocentric radius $r$.  We shall take $\alpha=2$ in this work when evaluating numerical results, but our approach will hold in general.  The mass function obeys the relation
\begin{equation}
     f\rho(r) = \int_{\Mmin}^{\Mmax}\! dM_h\, M_h
     \frac{dn_h}{dM_h}(r,M_h) \equiv \langle M_h \rangle n_h (r),
\label{eqn:mass function relation}
\end{equation}
where $\rho(r)$ is the density profile of the
Milky Way halo, $\Mmin$ and $\Mmax$ are the masses of the smallest and
largest subhalos, and in the last equality we define the mean mass
$\langle M_h \rangle$ as well as spatial number density $n_h(r)$
of subhalos.  From Eq.~(\ref{eqn:mass function relation}) and the
assumed shape of the mass function, we obtain
\begin{eqnarray}
     \frac{dn_h}{dM_h}(r,M_h) &=& \frac{f \rho(r)}{\ln (\Mmax/\Mmin)}
M_h^{-2},
\label{eqn:mass function}\\
n_h(r) &= & \frac{f\rho(r)}{\Mmin \ln(\Mmax/\Mmin)},
\label{eqn:microhalo density}\\
\langle M_h \rangle &=& \Mmin \ln(\Mmax / \Mmin),
\label{eqn:average microhalo mass}
\end{eqnarray}
where in Eq.~(\ref{eqn:microhalo density}), we assumed $\Mmin \ll \Mmax$.  We use the NFW \cite{NFW96} profile,
\begin{equation}
     \rho(r) = \frac{\rho_s}{(r/r_s) (1+r/r_s)^2},
\label{eqn:NFW}
\end{equation}
where $\rho_s = 5.4 \times 10^{-3}~M_{\odot}~\pcs^{-3}$
is the characteristic density, and $r_s = 21.7$ kpc is the scale
radius.  The density is set to zero beyond a cutoff radius $r_c = 10~r_s$, which is approximately the virial radius (i.e., the concentration parameter is $c \equiv r_{\mathrm{vir}}/r_s \approx 10$).
This normalizes the virial mass of the Milky Way halo to be $10^{12}
M_{\odot}$, and gives $\rho_0 = 7\times 10^{-3}~M_\odot~\pcs^{-3}$ as
the local density at the solar radius ($r_0 = 8.5$ kpc).

Following other studies \cite{Fornasa}, we normalize the mass function by using the results of simulations \cite{diemand} to fix the fraction of mass contained in high-mass microhalos.  Specifically, we choose $f$ such that $10\%$ of the total mass of the halo is contained in microhalos of mass $10^7 - 10^{10} M_\odot$.  We then extrapolate the power-law mass function found by the simulations down to a cutoff mass $\Mmin$ below the simulation resolution; taking $\Mmax = 10^{10} M_\odot$, $f$ then becomes a function of $\Mmin$:
\begin{equation}
    f(\Mmin) = 0.10\, \log(\Mmax/\Mmin)/\log(\Mmax/10^7 M_\odot).
\end{equation}
We shall suppress the argument $\Mmin$ when referring to $f$ below.

For a halo model with $\Mmin = M_{\oplus}$, we find that approximately $52\%$ of the total halo mass is contained in roughly $4.8\times 10^{16}$ microhalos in the specified mass range.  The number density of microhalos in the solar neighborhood is about $34\ \textrm{pc}^{-3}$.  In this paper, we will examine a class of halo models in which $\Mmin$ is a free parameter, and falls in the range $10^{-4}-10^{3}\,M_\oplus$ predicted by WIMP kinetic decoupling studies.

\subsection{Microhalo Annihilation Models}
\label{sec:annihilation}

Let us assume that the microhalos have NFW density profiles. The integrated number luminosity $L_h$ from WIMP annihilation in an microhalo with NFW profile parameters $r_s$, $c$, and $\rho_s$ is given by
\begin{equation}
    L_h = \frac{N_{\gamma}\langle\sigma v\rangle}{m_{\chi}^2} \int_h dV\, \rho^2
        \equiv a(c) K \rho_{s} M_h.
    \label{eqn:Lh}
\end{equation}
Here, $N_{\gamma}$ is the integrated number of photons per annihilating particle, $\langle\sigma v\rangle$ is the thermally-averaged annihilation cross-section multiplied by the relative velocity, and $m_{\chi}$ is the mass of the WIMP.  In the second equality,
\begin{equation}
    a(c) \equiv \frac{1-1/(1+c)^3}{3(\ln(1+c)-c/(1+c))}
\end{equation}
is a numerical factor resulting from the volume integral (with a dependence on $c$), and we have defined
\begin{equation}
    K \equiv \frac{N_{\gamma}\langle\sigma v\rangle}{m_{\chi}^2} = \frac{\langle\sigma v\rangle}{m_{\chi}^2} \int dE\, \frac{dN_{\gamma}}{dE}.
    \label{eqn:K}
\end{equation}
Here, $dN_{\gamma}/dE$ is the photon spectrum per annihilating particle. For the Galactic halo, using the NFW profile parameters defined in the previous section, we find that
\begin{equation}
    L_{MW} = 1.2\times 10^9 \, K\, M_{\odot}^2\, \pcs^{-3} = 5.1\times 10^{67}\, K\, \GeV^2\, \cms^{-3}.
    \label{eqn:LMW}
\end{equation}

We now also assume that the integrated gamma-ray number luminosity $L_h$ of each microhalo is proportional to its mass $M_h$, with constant mass-to-light ratio $\Upsilon \equiv M_h/L_h$.  Then, the luminosity function is $dn_h/dL_h = \Upsilon (dn_h/dM_h)$.  Note that throughout this paper, the luminosity is the {\it number} (not energy) of
photons emitted per unit time; similarly, we deal with number
fluxes (fluences) and intensities.

These assumptions are consistent with the results of simulations, which indeed roughly find that $L_h \propto M_h$.  In particular, Ref.~\cite{diemand} finds that
\begin{equation}
    \frac{L_h}{L_{MW}} = \frac{\int_h \rho^2 dV_h}{\int_{MW} \rho^2 dV_{MW}} \approx 3\times 10^{-12} \left(\frac{M_h}{M_{\odot}}\right)
\label{eqn:diemand ratio}
\end{equation}
in the range of their simulation, which resolves subhalos down to $M_h \approx 4\times10^6 M_{\odot}$.  We shall assume this relation holds down to the microhalo masses under discussion in this paper.  Note that Eq.~\ref{eqn:diemand ratio} essentially relates the microhalo NFW profile parameters $r_s$, $c$, and $\rho_s$ (which may be complicated functions of mass) to those of the Galactic halo, which were stated in the previous Subsection.

Combining Eqs.~\ref{eqn:Lh}-\ref{eqn:diemand ratio}, we can now parameterize the magnitude of the annihilation signal by the parameter $K$ (or equivalently, $\Upsilon^{-1} = 3.6\times 10^{-3}~K~M_{\odot}~\pcs^{-3} = 0.14~K~\GeV~\cms^{-3}$ or $\Lmin = \Upsilon^{-1} \Mmin$, which are both proportional to $K$).  Given that our halo model and microhalo mass function were parameterized by $\Mmin$, we see that our overall model has two parameters.  We now discuss a constraint on this model, arising from an intensity limit observed by EGRET.


\subsection{EGRET Constraints}
\label{sec:constraints}

The gamma-ray intensity $I_h(\psi)$ (units of
photons$~\cmsec~\sr^{-1}$) from microhalos along a line of sight at an angular separation $\psi$ from the Galactic center can be estimated as
\begin{eqnarray}
     I_h(\psi) &=& \frac{1}{4\pi} \int dl \int_{\Lmin}^{\Lmax}\!
     dL_h\, L_h \frac{dn_h}{dL_h}\left(r(l,\psi),L_h\right)
     \nonumber \\
     &=& \frac{f}{4 \pi \Upsilon} \int
      dl\, \rho\left(r(l,\psi) \right),
\label{eqn:intensity}
\end{eqnarray}
where $l$ is the distance along the line of sight; i.e.,
$r^2=r_0^2+l^2 -2 r_0 l \cos\psi$. Compare Eq.~\ref{eqn:intensity} with the intensity $I_G(\psi)$ from annihilation in the smooth component of the Galactic halo, which contains a fraction $1-f$ of the total halo mass:
\begin{equation}
    I_G(\psi) = \frac{K (1-f)^2}{4\pi} \int dl\, \rho^2\left(r(l,\psi) \right).
    \label{eqn:smooth}
\end{equation}
Note that $I_h$ and $I_G$ depend differently on $\rho\left(r(l,\psi) \right)$, causing them to vary differently with $\psi$.

Current upper bounds to the diffuse gamma-ray background from EGRET
place an upper limit on $I_h+I_G$.  However, because of the lower
energy range of EGRET, these upper limits apply to energies in the
range $0.1\,\GeV \leq E \leq 10\, \GeV$.  Fermi will be more sensitive
to photons with energies above $10~\GeV$ (due to larger volume and
better angular resolution at higher energies).  For any given
annihilation model, we are thus interested in the signal of gamma rays above $10~\GeV$, but must also check to see that the constraint in the lower energy range is obeyed.  We see that we must examine the energy dependence of $I_h+I_G$, and hence the annihilation photon spectrum $dN_{\gamma}/dE$, in order to properly apply these constraints.  We shall consider the annihilation photon spectrum in two different scenarios.

In the first scenario, we assume that the WIMP is a neutralino, resulting in a photon spectrum per annihilating particle fit by an analytic approximation given by Ref.~\cite{Bergstrom}:
\begin{equation}
    \frac{dN_{\gamma}}{dE} = \frac{1}{m_{\chi}} \frac{0.42 e^{-8x}}{x^{1.5}+0.00014},
    \label{eqn:spectrum}
\end{equation}
where $x \equiv E/m_{\chi}$.  For the neutralino particle properties, we choose typical values used in the literature.  We set $\langle\sigma v\rangle = 3\times 10^{-26}~\cms^3~\second^{-1}$, which reproduces the observed dark matter density if the WIMP is a thermal relic.  We also choose $m_{\chi} = 85~\GeV$; this choice maximizes $K$ for photon energies above $10~\GeV$, and hence maximizes the annihilation signal.

Along these lines, when discussing annihilation signals in this scenario, we shall redefine the parameter $K$ in all of the relevant preceding equations by using Eq.~\ref{eqn:spectrum} in Eq.~\ref{eqn:K}, and integrating only over the energy range of interest to Fermi ($E \geq 10\, \GeV$).  Definitions for $\Upsilon$, $L_h$, $I_h$, etc. in this energy range follow.  With these values, we find the annihilation parameter for the neutralino model
\begin{equation}
    K_N = 4.2\times 10^{28}\, \pcs^{3}\, \second^{-1}\, M_{\odot}^{-2} = 9.9\times 10^{-31}\,\cms^{3}\, \second^{-1}\, \GeV^{-2}.
    \label{eqn:Khigh}
\end{equation}
By choosing these properties, we fix the annihilation parameter $K$; our model then depends only on the single parameter $\Mmin$.  Hereafter, we shall refer to our overall model in this scenario as the ``neutralino model.''

To constrain this model, we rule out values of $\Mmin$ that result in intensities exceeding upper limits on the diffuse gamma-ray background found by EGRET (see Ref.~\cite{Sreekumar}).  That is, for a given $\Mmin$, we require $dI_{h}/dE~+~dI_{G}/dE~\leq~dI_{\mathrm{obs}}/dE$ over the EGRET energy range.  Ref.~\cite{Sreekumar} found that the gamma-ray background as observed by EGRET is roughly isotropic (after masking out the Galactic plane and center), and is suitably parameterized by $dI_{\mathrm{obs}}/dE~\approx~2.7~\times~10^{-8}~(E/6.5~ \GeV)^{-2.1}~\cmsec~\sr^{-1}~\GeV^{-1}$.  For our choice of neutralino properties, the relative shapes of the background and annihilation spectra are such that if the constraint $dI_{h}/dE+dI_{G}/dE \leq dI_{\mathrm{obs}}/dE$ holds at $6.5~\GeV$, then it is also satisfied over the entire energy range; thus, it suffices to check the constraint at this energy.  We find that the constraint is satisfied for all $\Mmin$ in the range $10^{-4}-10^{3}\,M_\oplus$ predicted by kinetic decoupling studies.

\begin{figure}[t]
\centering
\includegraphics{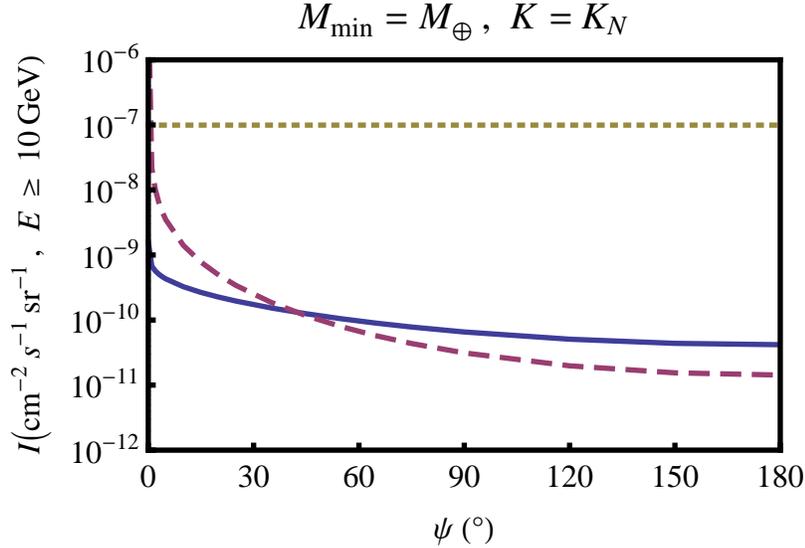}
\caption{The intensity $I_h(\psi)$ above $10~\GeV$ from microhalos as a
     function of the angle $\psi$ the line of sight makes with
     the Galactic center (solid) for the fiducial photon spectrum given by Eq.~\ref{eqn:spectrum} (the ``neutralino model'') and a cutoff mass of $\Mmin = M_{\oplus}$.  We also plot (dashed) the angular variation of the intensity $I_G(\psi)$ above $10~\GeV$ from dark-matter annihilation of a smooth component that contains $1-f$ of the total halo mass.  Note that the variation with $\psi$ of the gamma-ray flux from substructure is not as dramatic as that from annihilation in a smooth component.  The intensity $dI_{\mathrm{obs}}/dE$ integrated above $10~\GeV$ of the gamma-ray background as measured by EGRET (dotted) is also indicated.}
\label{fig:intensity}
\end{figure}

We plot in Fig.~\ref{fig:intensity} the angular dependence of the microhalo intensity $I_{h}$ above 10~GeV for a model in this scenario with fiducial cutoff mass $\Mmin = M_{\oplus}$.\footnote{Note that if $\rho(r)\propto r^{-1}$ as
$r\rightarrow 0$, the intensity is formally infinite at
$\psi=0$.  However, the flux from any finite-size window about the
Galactic center involves an integral over the intensity, and the
divergence of $I(\psi)$ at $\psi=0$ is such that the flux is
always finite.}  We also plot $I_G$, the angular dependence of the intensity above $10~\GeV$ from dark-matter annihilation from a smooth component containing $1-f$ of the total Galactic halo mass, in order to show that it varies more rapidly with $\psi$ than the angular dependence of the gamma-ray intensity from substructure.  We also show the intensity $dI_{\mathrm{obs}}/dE$ of the gamma-ray background as measured by EGRET, integrated above $10~\GeV$.

In the second scenario, we assume that WIMPs annihilate into monoenergetic gamma-rays of energy $E = 10~\GeV$.  We leave $K_E$ as a free parameter.
We approximate the upper limit from Ref.~\cite{Pullen:2006sy} to
the gamma-ray line intensity, averaged over the
$10^\circ\times10^\circ$ region around the Galactic center, by
$2\times10^{-6}\,(E/\GeV)^{-1/2}\,\cmsec~\sr^{-1}$ over the
energy range $0.1\,\GeV \leq E \leq 10\, \GeV$ (see also
Ref.~\cite{Mack:2008wu}), and we then derive an upper limit,
\begin{equation}
     f\Upsilon_E^{-1} \lesssim 10^{29}\, f_{I,h,l} \,
     M_\odot^{-1}\, \second^{-1} \approx 9 \times 10^{-29}\, f_{I,h,E} \, \GeV^{-1}\, \second^{-1}.
\label{eqn:LMupperlimit}
\end{equation}
Here, $\Upsilon_E^{-1} = 0.14~K_E~\GeV~\cms^{-3}$ is the light-to-mass ratio under the assumption of mono-energetic annihilation.  Also, $f_{I,h,E} \equiv I_{h,E}/(I_{h,E}+I_{G,E}+I_{d,E}) \leq 1$ is the fraction of the total gamma-ray intensity at $10~\GeV$ from the Galactic center arising from annihilation in microhalos, and depends on the residual intensity $I_{d,E}$ from any astrophysical backgrounds that may not have been subtracted in Ref.~\cite{Pullen:2006sy}.

In the case $I_{d,E}$ is negligible (i.e., the line intensity limit is saturated, with the observed intensity arising entirely from annihilation in the smooth halo and substructure), then $f_{I,h,E}$ depends only on the halo model and is a function of $\Mmin$.  Using Eqs.~\ref{eqn:intensity} and \ref{eqn:smooth}, calculation shows that a good estimate is given by $f_{I,h,E} \approx 0.086 \left(\Mmin/M_{\oplus}\right)^{-0.081}$.  For a given $\Mmin$, the intensity limit then provides an upper bound on $K_E$; using Eq.~~\ref{eqn:LMupperlimit} gives
\begin{equation} \label{eqn:KElimit}
K_E \lesssim 2\times 10^{-29} \frac{\left(\Mmin/M_{\oplus}\right)^{-0.81}}{f(\Mmin)} \,\cms^{3}\, \second^{-1}\, \GeV^{-2}.
\end{equation}
Thus, in this scenario we shall consider models parameterized by $\Mmin$ and $K_E$, constrained by this limit.  We shall refer to models in this scenario as ``line models.''  Comparing Eq.~\ref{eqn:KElimit} to Eq.~\ref{eqn:Khigh}, we note that these line models may have much larger fluxes than the neutralino model.

In Sections~\ref{sec:onepoint}-\ref{sec:genfeat}, we shall discuss the neutralino model, using the assumed form of the photon spectrum to predict the PDF for this fiducial model.  In Section~\ref{sec:detect}, we shall examine how observation of the PDF may place constraints on the $K_E-\Mmin$ parameter space for line models.

\section{Calculation of the PDF} \label{sec:onepoint}

The Fermi angular resolution at energies above $10~\GeV$ is roughly $0.1^\circ$; throughout this paper we shall assume square pixels of solid angle $(0.1^\circ)^2$.  This implies that the background flux will be measured in $\sim 4\times10^6$ beams on the sky.  One can then make a histogram of the number of counts in each beam.  Our goal here is to make predictions for the shape $P(F)$ for the distribution of these fluxes, under the assumption that these photons come from dark-matter annihilation in a clumpy Galactic halo.

Although $P(F)$ will in general be a function of the
line-of-sight direction $\psi$, we shall suppress this
dependence in much of the presentation,  reinserting it later
when required for numerical results.  We also
refer to all probability distribution functions as $P(x)$; the
particular function under discussion should be clear from the
argument $x$.

If the population of sources has a flux-density distribution
$P_1(F)$, then the probability $P(F)$ to see a total flux $F$
(integrated over all sources in the beam) in a given beam is
\cite{Scheuer:1957}
\begin{equation}
     P(F)  = \mathcal{F}^{-1} \!
     \left\{e^{\mu\left(\mathcal{F}\!\left\{P_1(F)\right\} -
     1\right)} \right\}.
\label{eqn:scheuersequation}
\end{equation}
Here $\mathcal{F}\{x\}$ is the Fourier transform of $x$ and
$\mathcal{F}^{-1}$ its inverse, and the flux-density distribution
$P_1(F)$ is normalized to $\int dF\, P_1(F)=1$.  The quantity
\begin{equation} \label{eqn:mu}
     \mu(\psi) = \frac{\Omega_{\mathrm{beam}} f}{\langle M_h\rangle}
     \int_0^{l_c(\psi)}\! dl'\, l^{\prime 2} \rho(r(l',\psi)),
\end{equation}
is the mean number of sources in each beam of solid angle
$\Omega_{\mathrm{beam}}$ (in sr).  We reproduce in the
Appendix the derivation of Eq.~(\ref{eqn:scheuersequation})
originally provided by Ref.~\cite{Scheuer:1957} (see also \cite{PD}).

\subsection{Derivation of $P_1(F)$} \label{subsec:P1F}

The first step is thus to find the flux-density distribution
$P_1(F)$ for individual sources in the beam.  This depends
on the luminosity function and on the spatial distribution of
microhalos.  The luminosity function is $P(L_h) \propto
L_h^{-\alpha}$, where again we take $\alpha=2$.  The probability for an individual microhalo to be at a distance $l$ along a line of sight $\psi$ is $P(l,\psi) \propto l^2 \rho(r(l,\psi))$.  We take a
maximum cutoff at $l_c(\psi)$, corresponding to a cutoff radius
$r_c =r(l_c(\psi),\psi)$.

\begin{figure}[b]
\centering
\includegraphics{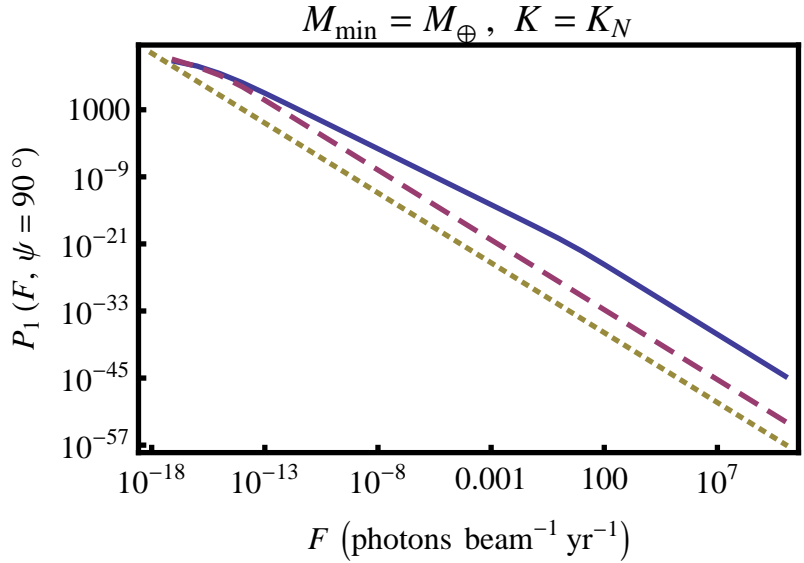}
\caption{The flux-density distribution
     $P_1(F,\psi=90^{\circ})$, normalized to unity, for the flux from an individual microhalo drawn from a population of microhalos with: (1) an NFW spatial distribution and a luminosity function
     $\propto L_h^{-2}$, for a fiducial value of the minimum cutoff luminosity $\Lmin$ (solid); (2) an NFW spatial distribution and uniform luminosity $\Lmin$ (dashed); and (3) a homogeneous spatial distribution and uniform luminosity $\Lmin$ (dotted).  Note that the first distribution follows a power-law of $F^{-2}$ in the probabilistically observable range of $F$ (following the power-law of the mass function), and then tends to $F^{-5/2}$ at extremely large $F$.  The last two distributions also tend to $F^{-5/2}$.  We have assumed the neutralino model and a cutoff mass $\Mmin = M_{\oplus}$.}
\label{fig:P1}
\end{figure}

We then find $P_1(F)$ is given by
\begin{eqnarray} \label{eqn:P1}
    P_1(F,\psi)
    &= \int dl\, dL_h\, P(l,\psi)P(L_h) \delta\left(F-\frac{L_h}{4\pi l^2}\right) \nonumber\\
    &\propto \int_0^{l_c(\psi)}\! dl\, l^4 \rho(r(l,\psi))\, (l^2 F)^{-\alpha} \theta\left(4\pi l^2 F-\Lmin\right) \theta\left(\Lmax - 4\pi l^2 F\right) \nonumber\\
    &\propto F^{-\alpha} \int_{l(\Lmin,F)}^{\min\left[l_c(\psi),l(\Lmax,F)\right]}\! dl\,
    l^{4-2\alpha} \rho(r(l,\psi)),
\end{eqnarray}
where the step functions enforce the cutoffs in $P(L_h)$, and $l(L_i,F)\equiv(L_i/4\pi F)^{1/2}$.  Note also the implicit cutoff in $P_1(F)$ for $F<\Lmin/4\pi l_c^2$.  Eq.~(\ref{eqn:P1}) can be evaluated numerically for a given value of the parameter $\Lmin = \Upsilon^{-1} \Mmin$.  The result is presented in Fig.~\ref{fig:P1} for the neutralino model, with the fiducial cutoff mass $\Mmin = M_{\oplus}$.

Note that Eq.~(\ref{eqn:P1}) yields the familiar
$P_1(F) \propto F^{-5/2}$ (conventionally written as
$N\left(>S\right) \propto S^{-3/2}$) for a homogeneous spatial distribution of sources with a general luminosity function, if the condition $l_c(\psi) \geq l(\Lmax,F)$ is satisfied over the range of $F$ of interest.  Under this condition, $P_1(F)$ will also asymptote to $F^{-5/2}$ at large $F$ for a non-pathological spatial distribution.  However, if $l_c(\psi) < l(\Lmax,F)$ for values of $F$ within the range of interest, then there will be a break in $P_1(F)$; $P_1(F)$ will tend to $F^{-\alpha}$ at $F$ for which the second condition holds, and will then tend to $F^{-5/2}$ at higher $F$.

For the problem under discussion, values of $\Lmax$ in the interesting regions of parameter space are such that $P_1(F)$ is negligible in the $F^{-5/2}$ regime.  Thus, the essential ``large-$F$'' dependence of  $P_1(F)$ will be $F^{-\alpha}$.

\subsection{Calculation of the Counts Distribution}
\label{subsec:PF}

The function $P(F)$ gives the probability to observe a flux $F$ from annihilation in substructure.  Unlike the function $P(F_{\mathrm{BG}})$ giving the probability to observe a flux $F_{\mathrm{BG}}$ from smoothly-distributed background sources (such as annihilation in the smooth component of the Galactic halo or other diffuse backgrounds), $P(F)$ will not be a Poisson distribution.

We will observe these fluxes in terms of
$\mathrm{photons}~\mathrm{beam}^{-1}~\mathrm{year}^{-1}$, or
similar units.  However, the limits from EGRET require that the mean photon count per beam per year be less than one; thus, even those beams with the highest photon counts will only observe some small integer number of photons per year.  It follows that we will need to discretize the continuous variables $F$ and $F_{\mathrm{BG}}$. Furthermore, emission of photons is a Poisson process.  Thus, let the total number of photons measured in a given beam over an observation period $T$ be $C \approx E(F+F_{\mathrm{BG}}) \in\mathbb{N}$; here $E$ is the exposure in a beam given in units of $\mathrm{cm}^2~\mathrm{sec}$, and is given by $E \approx AT$, where $A \approx 2000\, \mathrm{cm}^2$ is the area of the detector.

The discrete probability distribution $P(C)$ is then given by the sum of Poisson distributions with mean $E(F+F_{\mathrm{BG}})$ weighted by $P(F)$:
\begin{equation} \label{eqn:PC}
     P(C) = \int_0^{\infty}\! dF\, P(F) \wp(E(F+F_{\mathrm{BG}}),C),\qquad C \in \mathbb{N}.
\end{equation}
The shape of the discrete distribution $P(C)$ is generally very
similar to that of the continuous distribution $P(EF)$ and is
only slightly modified at the low end.

\begin{figure}[p]
\centering
\includegraphics{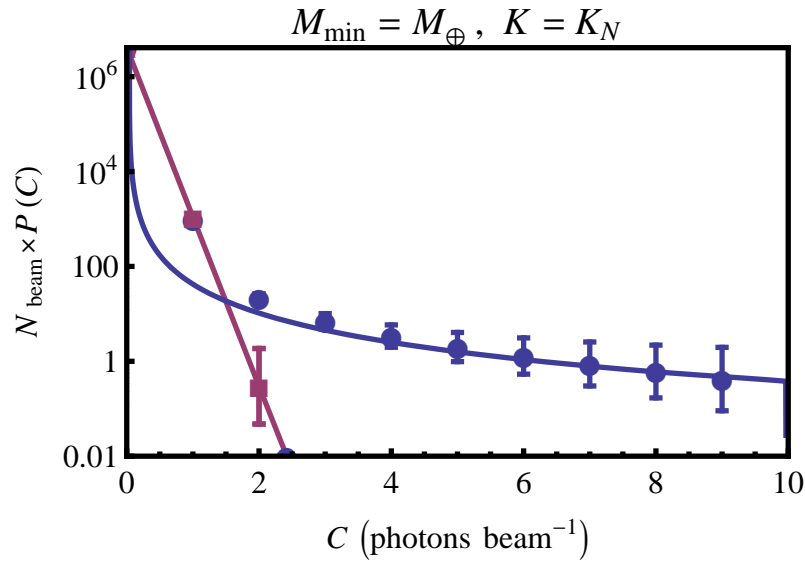}
\caption{The angular-averaged discrete probability distribution function
     $P(C)$ for the total photon number $C$ in a given
     beam (circles), for the neutralino model with a cutoff mass of $\Mmin = M_{\oplus}$.  Angular bins with widths of $\Delta\psi = 20^{\circ}$ were used in the averaging.  Only counts from annihilation in substructure and the smooth halo component have been included.  An observation period of 10 years has been assumed.  The angular-averaged continuous $P(F)$ (solid) and a fitted Poisson distribution (squares) are also plotted for comparison.  We have normalized to $N_{\mathrm{beam}}=4\pi / (0.1^{\circ})^2 \approx 4\times10^6$, the number of beams at the angular resolution limit.  Poisson error bars are also shown.}
\label{fig:PC}
\end{figure}

\begin{figure}[p]
\centering
\includegraphics{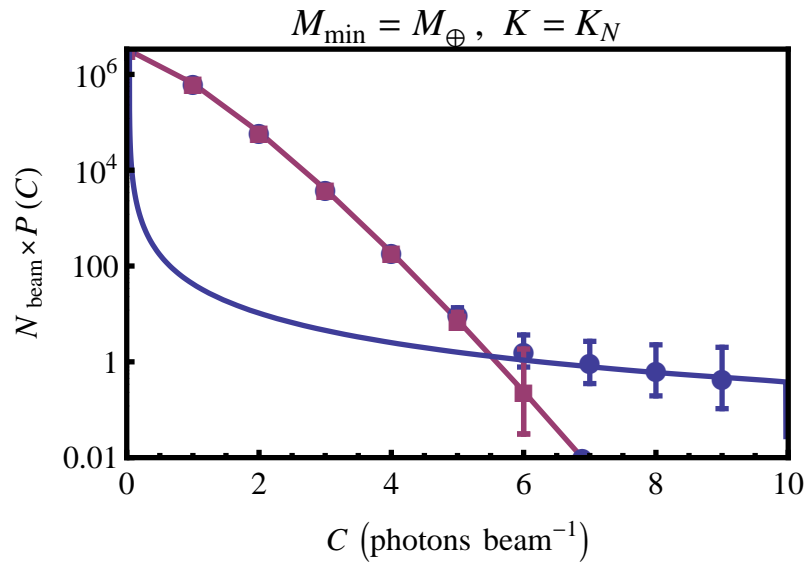}
\caption{The same as in Fig.~\ref{fig:PC}, but with an arbitrary diffuse background with intensity $I_d = 10^{-7}~\cmsec~\sr^{-1}$ above $10~\GeV$ added.  This additional background adds a large Poisson-like feature to $P(C)$ at low $C$, which obscures the substructure power-law tail.  This suggests that the neutralino model may be just outside the range of P(D) analysis, if the diffuse background is indeed this large.}
\label{fig:PCBG}
\end{figure}

\begin{figure}[t]
\centering
\includegraphics{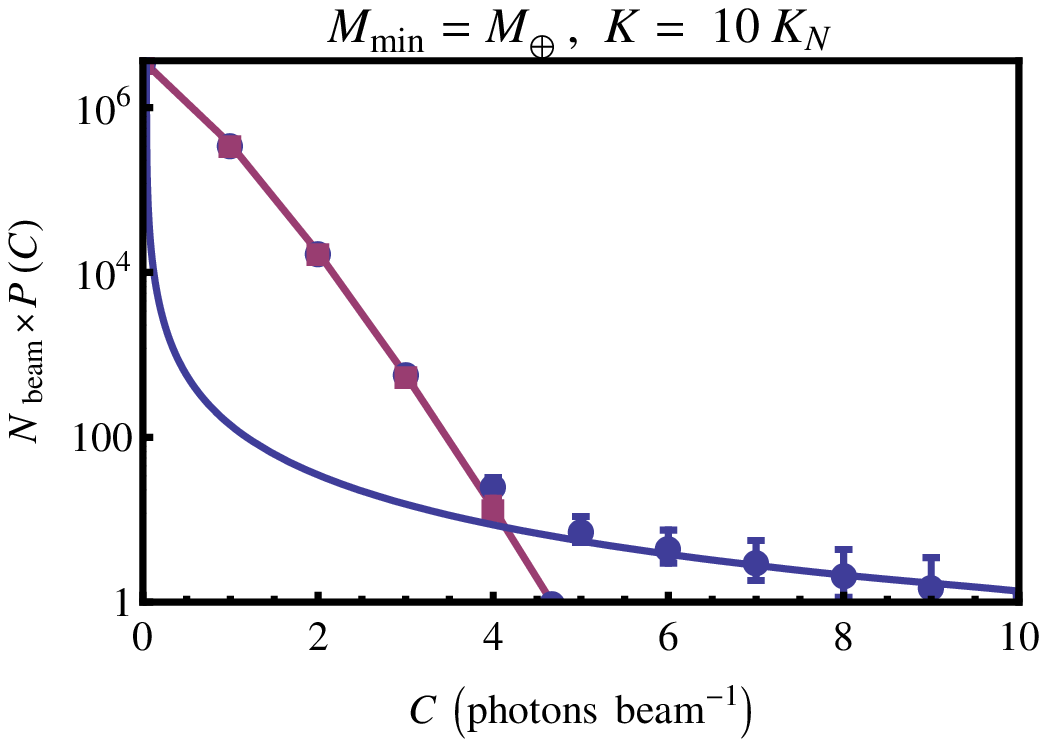}
\caption{The same as in Fig.~\ref{fig:PCBG}, but for a line model with $K_E=10\,K_N$, an observation period of five years, and a background intensity $I_{d,E} = 10^{-7}~\cmsec~\sr^{-1}$ at $10~\GeV$.  Note that although the intensity from the diffuse background is still many orders of magnitude above the mean expected intensity from annihilation in substructure, the substructure power-law tail is detectable with high statistical significance.  Furthermore, for these model parameters, there will be no detectable individual microhalos.  Thus, P(D) analysis may be useful in detecting substructure even when individual point sources are not detected.}
\label{fig:PCLine}
\end{figure}

\section{Numerical Results} \label{sec:genfeat}

Fig.~\ref{fig:PC} shows the results of numerical tabulation of the PDF $P(F)$ for the neutralino model with a fiducial cutoff mass of $\Mmin = M_{\oplus}$.  The PDF has a peak at low $F$ and a power-law tail at high $F$. Here, only flux from dark matter annihilation in substructure and the smooth halo are included.  Fig.~\ref{fig:PCBG} adds an additional diffuse background with intensity $I_d = 10^{-7}~\cmsec~\sr^{-1}$ above $10~\GeV$; in this case, flux from annihilation only comprises a small fraction of the total observed flux, and the substructure PDF may be difficult for Fermi to detect within the mission lifetime.

If the mass-to-light ratio $\Upsilon$ is increased (equivalently, if $K$ is decreased), with $\Mmin$ held fixed, then the photon flux decreases.  The entire distribution is then scaled down along the $F$-axis.  If $\Mmin$ is reduced, with $\Upsilon$ held fixed, then the relative width of the peak of the PDF decreases. This behavior can be understood by considering the limit $\Mmin \rightarrow 0$; in this case, we should expect to recover a smooth spatial distribution, resulting in a delta-function $P(F)$.  Note that this dependence on $\Mmin$ implies that the peak of the PDF must be resolved in order to measure $\Mmin$; if the peak is obscured by an extraneous diffuse background, as in Fig.~\ref{fig:PCBG}, then $\Mmin$ may be a degenerate parameter.

The distribution $P(C)$ for discretized counts $C$ is also
plotted in Figs.~\ref{fig:PC}~and~\ref{fig:PCBG}, for a ten-year Fermi exposure.  Also plotted is the Poisson distribution for a smoothly-distributed diffuse background of the same mean flux.  As the Figures indicate, the large-$F$ power-law tail of the PDF is
qualitatively different than the exponential falloff of the
Poisson distribution with $F$.  Thus, detection of substructure
amounts to detection of such a power-law tail.

Furthermore, the power-law tail of $P(F)$ follows the power-law tail of $P_1(F)$.  This is simply because single bright sources dominate beams with high $F$. However, as discussed in Subsection~\ref{subsec:P1F}, the power-law tail of $P_1(F)$ in turn follows the power-law of the mass function.  For example, the power-law tail in Fig.~\ref{fig:PC} indeed follows an $F^{-2}$ dependence.  Thus, $P(F)$ not only provides a method of substructure detection; it can also reveal the substructure mass function.

\section{Detectability}\label{sec:detect}

Figs.~\ref{fig:PC}~and~\ref{fig:PCBG} are plotted for the neutralino model in which a fiducial value of $K$ is chosen.  This model predicts a mean flux far below the EGRET continuum limit; even Fermi may have to observe for a period of at least ten years in order to detect significant numbers of photons in beams in the substructure power-law tail.  However, the constraint on line models given by Eq.~\ref{eqn:LMupperlimit} allows for choices of the parameters $K_E$ and $\Mmin$ that result in mean fluxes much closer to the EGRET line intensity limit.  Line models that saturate the limit will produce signals that could be easily detected by Fermi within a year.  Fig.~\ref{fig:PCLine} plots the PDF for a fiducial line model.

Of course, Fermi will also be sensitive to a range of line models predicting fluxes below the EGRET bound.  However, for line models with mean fluxes below a certain level, the amplitude of the substructure power-law tail will so reduced that it will be impossible to detect, as it was for the neutralino model in Fig.~\ref{fig:PCBG}.  In this Section, we determine the regions of the line model parameter space in which the PDF can be distinguished from the Poisson distribution expected for a completely smooth or diffuse background of the same mean flux, over an observation period of five years.  Combined with the EGRET limit, this analysis will show the region of allowed parameter space that can be probed by study of the PDF.

\begin{figure}[t]
\centering
\includegraphics{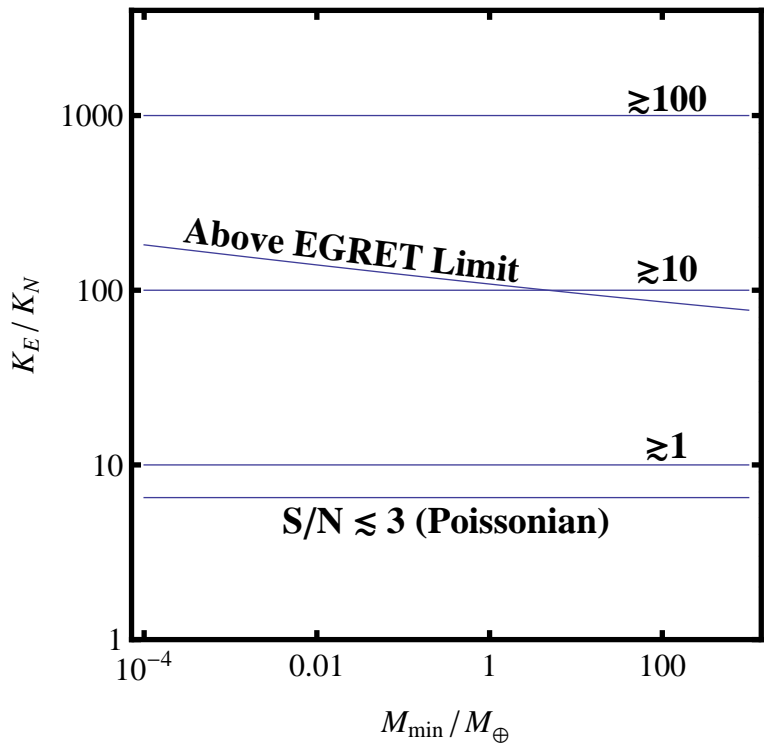}
\caption{The $K_E$-$\Mmin$ parameter space for the line models.  On the
     vertical axis, $K_E$ is scaled by $K_N$, the annihilation parameter for the neutralino model given by Eq.~\ref{eqn:Khigh}.  We
     indicate the region that is already ruled out by the
     EGRET upper limit to the diffuse background; where
     there will be $\gtrsim 1$, $\gtrsim 10$, and $\gtrsim 100$ detectable point sources with flux greater than the five-year point-source sensitivity of Fermi; and where measurements of the flux PDF cannot be distinguished from a Poisson distribution, for an observation period of five years.  Angular bins with widths of $\Delta\psi = 20^{\circ}$ were used in the calculation of S/N, and regions near the Galactic center ($\psi \leq 30^{\circ}$) were masked.}
\label{fig:parameterspace}
\end{figure}

We determine the signal-to-noise with which the PDF
$P(C)$ can be distinguished from the Poisson distribution
$\wp\left(\VEV{C},C \right)$ with the same mean count rate $\VEV{C}$.
The null hypothesis of no substructure can be eliminated at the
$3\sigma$ level if $S/N>3$, where
\begin{equation} \label{eqn:StoN}
     \frac{S}{N} = \sqrt{\sum_{\psi_i} N_{\mathrm{beam,bin}}(\psi_i) \left(\frac{S}{N}\right)_{\psi_i}^2},
\end{equation}
and
\begin{equation}
    \left(\frac{S}{N}\right)_{\psi_i}^2 = \sum_{C=0}^{C_{\mathrm{max}}(\psi_i)}
     \frac{\left[P(C,\psi_i)-\wp\left(\VEV{C}_{\psi_i},C \right)
     \right]^2}{ \wp\left(\VEV{C}_{\psi_i},C\right)}.
\end{equation}
Here, we label the angular bins by the central value of the bin $\psi_i$. The quantity $N_{\mathrm{beam,bin}}(\psi_i)$ is the number of beams contained in each bin, $C_{\mathrm{max}}(\psi_i)$ is the highest count observed in each bin, and $\VEV{C}_{\psi_i}$ is the mean of the best-fit Poisson distribution in each bin.  Eq.~(\ref{eqn:StoN}) then quantifies the difference between the discrete probability distributions $P(C,\psi)$ and $\wp\left(\VEV{C}_{\psi},C \right)$, comparing the substructure PDF with the Poissonian distribution expected from a diffuse background (which may have angular dependence). In Fig.~\ref{fig:parameterspace}, we plot the regions of the $K_E$-$\Mmin$ parameter space in which the value of $S/N$ indicates that substructure can be detected.  Also plotted are the regions of the parameter space ruled out already by the current EGRET upper limit to the diffuse background.

Note that in Fig.~\ref{fig:parameterspace}, models fall in the Poissonian regime when the substructure power-law tail at high $C$ is obscured by the Poisson-like feature at low $C$; in this regime, P(D) analysis cannot be used.  Since the Poisson-like feature arises from the flux from extraneous diffuse backgrounds (from sources other than dark matter annihilation in substructure and the smooth halo), the demarcation of the Poissonian regime is ultimately determined by the level of these backgrounds.  In determining the Poissonian regime, we have conservatively assumed an arbitrary diffuse background of $I_{d,E} = 10^{-7}~\cmsec~\sr^{-1}$; in this case, the EGRET intensity limit is severely unsaturated, and only a small fraction of the observed diffuse background arises from dark matter annihilation in substructure and the smooth halo.  In practice, the actual level of these diffuse backgrounds will determine the Poissonian regime, which may then cover a smaller region of parameter space than the conservative estimate presented in Fig.~\ref{fig:parameterspace}.

Of course, detection of a nontrivial PDF is also intimately related
to the criteria for detection of point sources.  The number of sources observed with flux greater than $F$ is given by
\begin{eqnarray} \label{eqn:detect}
N(\geq F) &= 2\pi \int d\psi\, \sin\psi \int dL_h \int_0^{l(L_h,F)}\! dl\, l^2 \frac{dn_h}{dL_h}(r(l,\psi),L_h).
\end{eqnarray}
Examining this equation shows that $N(\geq F)$ is only weakly dependent on the cutoff mass $\Mmin$.  Furthermore, since the observed microhalos essentially comprise a volume-limited sample, for $K_E$ in the range of interest $N(\geq F) \propto K_E$ (at lower $K_E$, $N(\geq F) \propto K_E^{3/2}$ as expected for a flux-limited sample).  Numerical calculation of $dN(\geq F)/dM$ shows that the observed microhalos will predominantly be those of higher mass $\gtrsim 10^3~M_{\odot}$; although lower-mass microhalos are far more numerous, Fermi will not be sensitive enough to detect them individually~\cite{proper2}.

In certain regions of the parameter space for which
Fig.~\ref{fig:parameterspace} indicates a nontrivial PDF,
substructure will be detectable via detection of individual
microhalos, even without a detailed analysis of the PDF.  We
plot these regions, taking the Fermi five-year $5\sigma$ point-source sensitivity at $10~\GeV$ of $F \approx 2\times 10^{-10}~\cmsec$~\cite{glast} (note that this sensitivity assumes the same background level as in our determination of the Poissonian regime).  The advantage of the full PDF, however, is that substructure can be detected even in regions of parameter space where individual microhalos elude detection.  Measurement of the detailed shape of the PDF can also provide more information on the microhalo mass function and/or spatial distribution in the halo than would be obtained
simply by point-source counts; e.g., the slope of the power-law
tail in the PDF depends upon the slope of the mass function.

Note that we could have done a similar analysis for a more general WIMP and substructure model in which some of the parameters (e.g., annihilation cross-section and spectrum, or subhalo concentration/boost parameters) were allowed to vary.  The EGRET continuum constraint allows for a large range of such models.  However, for simplicity we have only considered the parameter space of line models and the specific substructure model assumed in Section~\ref{sec:egret}.

\section{Conclusions and Comments}
\label{sec:conclusions}

We have proposed that the distribution of fluxes measured
in individual Fermi pixels can be used to probe the existence of
substructure in the Galactic halo to very small mass scales.
By characterizing fluctuations in the diffuse gamma-ray
background in this way, the existence of Galactic substructure
may be inferred statistically even if individual halos cannot be
detected.  This statistical approach should be viewed as
complementary to the use of an angular correlation function
\cite{anisotropy1, anisotropy3}.  Since the PDF is a convolution of
the microhalo mass function and spatial distribution,
constraints to the parameters of these distributions may be
obtained by measuring the PDF.

The full PDF we have calculated may be useful even in situations
where individual microhalos can be detected.  For example, the
flux in a pixel with a $3\sigma$ excess which is interpreted as
detection of a single point source may actually be due to
several point sources; the probability that this is so may be
inferred from the PDF.

We have illustrated the PDF that results in a phenomenological
model for substructure parameterized a microhalo mass cutoff $\Mmin$, and a mass-to-light ratio $\Upsilon$.  This is almost certainly an
oversimplification.  In more realistic models,
the mass function may differ from the particular power law we
have assumed.  The mass-to-light ratio may depend on the
microhalo mass, and there may even be a spread of luminosities
for each mass.  The spatial distribution of microhalos may not trace the Galactic halo.  Similarly, contributions to the PDF from astrophysical backgrounds (e.g., from cosmic-ray spallation or extragalactic sources) may need to be considered before a complete comparison of our model predictions with data can be made \cite{Dodelson}.

In our P(D) analysis, we did not consider the dependence of the angular resolution on the photon energy. Furthermore, we have also assumed here that each microhalo will fall within a single resolution element of Fermi.  Taking into account the finite angular size of each microhalo will reduce the length of the power-law tails in the PDF, and will decrease the region of parameter space in which the PDF can probe substructure.  However, note that individual extended sources will also be more difficult to detect than point sources.  A generalization of Eq.~\ref{eqn:detect} will give a smaller number of detectable extended sources; the corresponding lines in Fig.~\ref{fig:parameterspace} will also shift upwards.  Thus, there will still be an appreciable region of parameter space in which the PDF can be used to detect substructure even if individual sources cannot be detected.  Moreover, a conservative rough estimate of the size of these microhalos can be found by approximating the microhalo mass density $\rho_h$, assuming a formation redshift of $z \approx 100$ and a concentration parameter of $c \approx 1$ \cite{earthmass}.  A simple calculation then gives the angular size of the closest and most extended microhalos as $\theta \approx \left(f \rho_0/\rho_h\right)^{1/3} \approx 4^{\circ} f^{1/3}$.  Thus, if the beam size is increased such that the majority of extended microhalos fall within a single beam, then the point source P(D) formalism presented here is roughly valid.  A more careful generalization may be required for comparison to data.

We leave the inclusion of these additional levels of complication
to future work.  In addition to these future directions, one may also consider going further by combining the angular-correlation and PDF
approaches.  For example, the full two-point flux probability
distribution function can be calculated and may provide
additional observables with which to constrain the models or to
distinguish a dark-matter background from other astrophysical
backgrounds.  Again, this is left for future investigation.

\ack
This work was supported by the Sherman Fairchild Foundation
(SA), DoE DE-FG03-92-ER40701, and the Gordon and Betty Moore
Foundation.

\section*{Appendix: Derivation of $P(F)$}\label{appendx:PF}

Here we derive the relation between the flux-density
distribution $P_1(F)$ and the flux PDF $P(F)$.
Such a calculation is termed ``$P(D)$ analysis'' in
the literature, as it was first performed for observations of
faint radio sources that produced ``deflections'' of the
measuring apparatus.  This $P(D)$ analysis is
useful in determining if an observed diffuse background is
actually composed of numerous faint point sources.  If this is
the case, then there will be fluctuations in the diffuse signal
from the random Poisson clustering of point sources in each
beam.  The shape of $P(F)$ thus depends not only on $P_1(F)$,
but also the mean number $\mu$ of sources [Eq.~(\ref{eqn:mu})]
in each beam.

We wish to find the probability distribution for a total flux
$F$ in a beam, given that it is the sum $F=\sum_{i}^k F_i$ of the
fluxes $F_i$ from individual microhalos.  Each of the
$F_i$ is a random variable with probability distribution
$P_1(F_i)$.  Furthermore, the number $k$ of fluxes $F_i$
entering into the sum is itself a random variable given by a
Poisson distribution with mean $\mu$. Let us call $P_k(F)$ the
probability that $k$ random variables $F_i$ sum to $F$; i.e.,
the probability that $k$ microhalos emit a total flux $F$.  Then
\begin{equation} \label{eqn:PFPoisSum}
     P(F) = \sum_{k=0}^{\infty} \wp\left(\mu,k\right) P_k(F),
\end{equation}
where $\wp\left(\mu,k\right)$ is a Poisson probability
distribution for $k$ with mean $\mu$.

It now remains to determine $P_k(F)$.  For $k=0$, it is clear
that $P_0(F) = \delta(F)$; $P_1(F)$ is given.  For $k>1$,
$P_k(F)$ is given by
\begin{equation} \label{eqn:PkF}
     P_k(F) = \int_0^{\infty}\!dF_1 \ldots \int_0^{\infty}\!dF_k\,
     \left(\prod_{i=1}^k P_1\left(F_i\right)\right)\, \delta(F -
     \sum_{i=1}^k F_i).
\end{equation}

The easiest way to compute Eq.~(\ref{eqn:PkF}) is to note that
the Dirac delta function transforms the integral into a
convolution \cite{Scheuer:1957,Willis:1996}.  To see this, let
us examine the integral for $k=2$:
\begin{eqnarray} \label{eqn:ConvDemo}
     P_2(F) &= \int_0^{\infty}\!dF_1 \int_0^{\infty}\!dF_2\,
     P_1(F_1)\, P_1(F_2)\, \delta(F-(F_1+F_2)) \nonumber\\
       &= \int_0^{\infty}\!dF_1\, P_1(F_1)\, P_1(F-F_1) \nonumber\\
       &= (P_1 \ast P_1)(F).
\end{eqnarray}
It follows that $P_k(F) = (P_{k-1} \ast P_1)(F)$; then by
induction, $P_k(F)$ is given by $P_1(F)$ convolved (or
autocorrelated) with itself $k$ times.  Using the convolution
theorem, it then follows that
\begin{equation} \label{eqn:PkFFT}
     P_k(F) =
     \mathcal{F}^{-1} \!
     \left\{\mathcal{F}\!\left\{P_1(F)\right\}^k\right\},
\end{equation}
where $\mathcal{F}$ denotes a Fourier transform.  Note that
Eq.~(\ref{eqn:PkFFT}) also holds for $k=0$ (and trivially for
$k=1$).

Inserting Eq.~(\ref{eqn:PkFFT}) into Eq.~(\ref{eqn:PFPoisSum})
and using the linearity of the inverse Fourier transform, we
find
\begin{eqnarray} \label{eqn:PF}
P(F) &= \sum_{k=0}^{\infty} \frac{e^{-\mu}\mu^k}{k!} \mathcal{F}^{-1}\!\left\{\mathcal{F}\!\left\{P_1(F)\right\}^k\right\}\nonumber\\
     &= e^{-\mu}\mathcal{F}^{-1}\!\left\{ \sum_{k=0}^{\infty} \frac{\left(\mu \mathcal{F}\!\left\{P_1(F)\right\}\right)^k}{k!}\right\}\nonumber\\
     &= \mathcal{F}^{-1}\!\left\{e^{\mu\left(\mathcal{F}\!\left\{P_1(F)\right\} - 1\right)} \right\}.
\end{eqnarray}
Eq.~(\ref{eqn:PF}) gives the desired relation for $P(F)$ in
terms of $P_1(F)$ and $\mu$.  Although the presence of the inverse
Fourier transform prevents further analytic simplification in
general, this expression can be computed numerically using fast
Fourier transforms on a discretized $P_1(F)$.

\end{document}